\documentclass[final,hidelinks,onefignum,onetabnum]{siamart250211}

\usepackage{amsfonts}
\usepackage{subfig}
\usepackage{amsmath}
\usepackage{amssymb}
\usepackage{booktabs}
\usepackage{bm}
\usepackage{hyperref}
\usepackage{tcolorbox}
\usepackage{natbib}

\hypersetup{colorlinks,citecolor=blue}

\newtheorem{identity}{Identity}
	
\newcommand*\diff{\mathop{}\!\kern0pt\mathrm{d}}

\newcommand{\Var}{\mathrm{Var}}
\newcommand{\Error}{\mathrm{Error}}
\newcommand{\Black}{\mathrm{Black}}

\ifpdf
\hypersetup{
	pdftitle={Asian Basket Spread Options: A New Approximation Based on Stochastic Taylor Expansions},
	pdfauthor={Fabien Le Floc'h}
}
\fi
\title{Asian Basket Spread Options: A New Approximation Based on Stochastic Taylor Expansions\thanks{Date: April 23, 2025.}}
\author{Fabien Le Floc'h\\ \email{fabien@2ipi.com}}
\date{April 23, 2025}

\begin{document}
	\maketitle
	\begin{abstract}
We present closed analytical approximations for the pricing of Asian basket spread options under the Black-Scholes model. The formulae are obtained by using a stochastic Taylor expansion around a log-normal proxy model and are found to be highly accurate for Asian and spread options in practice. Unlike other approaches, they do not require any numerical integration or root solving.
	\end{abstract}
	\section{Introduction}
	
	The buyer of a vanilla basket spread option of strike $K$ and maturity $T$ on  underlying assets $(S_i)_{i=1,...,m}$  receives at maturity $T$ the payoff \begin{equation}V(T)=\max\left(\sum_{i=1}^m w_i S_i(T) - K,0\right)\end{equation} for a call and $\max\left(K - \sum_{i=1}^m w_i S_i(T),0 \right)$ for a put. While for a vanilla basket option the weights $w_i$ are positive and the strike price obeys $K > 0$, for a spread option the weights $w_i$ are negative for  $i=1,...,m_p-1$, positive for $i=m_p,...,n$ and are usually chosen such that
	$\sum_{i=1}^{m_p-1} w_i = -1$ and $\sum_{i=m_p}^n w_i = 1$. The strike price $K$ may be positive or negative.
	Those are not so common in the domain of foreign exchange or equity derivatives (with $m>2$). The most popular example, the so-called \emph{crack} spread stems from commodity derivatives. The portfolio of the 3:2:1 crack spread reads $3S_3 - 2S_2 - S_1$, where $S_3, S_2$ and $S_1$ are respectively the prices of crude oil, gasoline and heating oil. Other commodities such as soybean or cattle may also involve spreads on three different future contracts. 
	
	The buyer of an Asian call spread option of strike $K$ and averaging dates $(t_j)_{j=1,...,n}$ on an asset $S$ receives \begin{equation}V(T)=\max\left(\sum_{j=1}^n w_j S(t_j) - K, 0\right)\,.\end{equation}
	Those are sometimes traded in equity and foreign exchange spaces.  Under the Black-Scholes model with a term-structure of interest rates, dividend yields and volatilities, Asian spread options with discrete arithmetic averaging can be reduced to an equivalent vanilla basket spread option with a specific correlation structure. 

Asian basket spread options combine both features: each underlying asset is averaged, typically over the same schedule and with the same weights, the payoff at maturity reads \begin{equation}
V(T) =\max\left(\sum_{i=1}^m \sum_{j=1}^nw^B_{i}  w^A_{j} S_i(t_j) - K,0\right)\,,\end{equation} 
where $\bm{w}^A$ are positive Asianing weights and typically sum to 1, while $\bm{w}^B$ are basket weights of mixed sign.
	Asian basket spread options are mostly traded by commodity traders.
	
	In the context of the Black-Scholes model, an early and popular approximation for prices of vanilla spread options (for $m=2$), is due to \cite{kirk1995correlation}. There exists however a simple semi-analytical exact formula based on a one-dimensional integral \citep{pearson1999efficient}, which should be preferred nowadays. As the number of assets $m$ (or the number of Asianing dates) increases, the integral formulation becomes less practical. \cite{castellacci2003asian} combined the approximation of the arithmetic mean by a geometric mean (originally used by \cite{vorst1992prices} for vanilla Asian options) with the exchange option formula of \cite{margrabe1978value} to obtain a simple approximation for the prices of Asian basket spread options. \cite{deelstra2010pricing} found approximations based on comonotonic bounds, which necessitate a one dimensional integration but compare favorably to the three moment matching shifted lognormal approximation of \cite{borovkova2007closed}. More recently, \cite{pellegrino2016general} derived more accurate approximations, which may be seen as an extension of \cite{bjerksund2014closed}.
	
	Here, we consider a Black-Scholes setting with time-dependent parameters, and propose a stochastic Taylor expansion using Margrabe exchange option formula on the geometric averages as a proxy, following the ideas of \cite{etore2012stochastic}, who applied their expansion to price vanilla options with discrete dividends under the piecewise-lognormal model. It may be seen as a generalization of our previous work regarding vanilla basket and Asian options in \citep{lefloc2024stochastic}. The resulting formula does not involve any numerical integration or root solving, thus offering computational speed and facilitating hedges calculations (for example using the option greeks). We start by deriving the proxy, the first-, second- and third-order expansions and then evaluate their accuracy on a sample of previously published cases. This allows for a comparison with other accurate approximations.

\section{Reductions to vanilla basket spread options}
\subsection{From Asian to basket}
Let $F(0,t)$ be the forward price of the underlying asset from valuation time to the date $t$. In the Black-Scholes model with time-dependent parameters we have \begin{equation*}F(0,t) = S(0)e^{\int_{0}^{t} r(s) - q(s) \diff s}\,,\end{equation*} with $r, q$ being the instantateneous interest rate and dividend yield and $S(t)$ the underlying asset spot at time $t$.  Under the risk-neutral measure $\mathbb{Q}$, the forward price follows
\begin{equation*}
	\diff F(t,T) = F(t,T) \sigma(t) \diff W(t)\,,
\end{equation*}
where $W$ is a $\mathbb{Q}$-Brownian motion.

The buyer of an arithmetic Asian option of strike $K$ and maturity $T$ on an underlying $S$ with observation dates $(t_i)_{i=1,...,n}$ with $t_n=T$, receives at maturity $\max\left(\sum_{i=1}^n w_i S(t_i) - K\right)$ for a call and $\max\left(K - \sum_{i=1}^n w_i S(t_i) \right)$ for a put. It may be expressed as a basket option of strike $K$ and maturity $T$ on $n$ underlying assets $(S_i)_{i=1,...,n}$. The covariance between the assets must match the covariance of $S$ between different observation times, this means\footnote{We use the It\^o isometry to obtain Equations \ref{eqn:correlation_asian} and \ref{eqn:correlation_asian_basket}.}
\begin{equation}
	\rho_{i,j} \sqrt{v_i v_j} =v_{i \wedge j}\,, \label{eqn:correlation_asian}
\end{equation}
where $\rho_{i,j}$ denotes the correlation between the assets $S_i$ and $S_j$, $v_i = \int_{0}^{t_i} \sigma^2 (s) \diff s$ being the total variance up to time $t_i$ for the asset $S$. 
Each underlying asset forward to $T$ must match the forwards of the single asset to $t_i$, meaning 
\begin{equation}F_i(0,T) = \mathbb{E}[S_i(T)] = \mathbb{E}[S(t_i)] = F(0,t_i)  = S(0)e^{\int_0^{t_i} (r(s)-q(s)) \diff s}\,.\end{equation}

Thus, the arithmetic Asian option on $S$ with observations $t_i$ corresponds to a vanilla basket option on $S_i$ with forwards to maturity $F_i(0,T)$, total variances $v_i$ and correlations $\rho_{i,j}$ for $i,j=1,...,n$.

\subsection{Asian baskets}
This may be extended to Asian baskets with payoff at maturity \begin{equation*}\max\left(\sum_{i=1}^{n} w_i \sum_{j=1}^{m}  \mu_j S_j(t_i)- K,0\right)\,,\end{equation*} where $(\mu_j)_{j=1...m}$ are basket weights and $S_j$ are the different assets with correlation $\rho_{j,l}$. We may define $n\times m$ assets $\hat{S}_{i+n(j-1)}$ with forward $F_j(t_i)$, total variance $\int_{0}^{t_i} \sigma_j^2(s) \diff s$, correlation
\begin{equation}\hat{\rho}_{i+n(j-1),k+n(l-1)} = \rho_{j,l} \frac{\int_0^{t_{i} \wedge t_k} \sigma_j(s) \sigma_l(s)\diff s}{\sqrt{ \int_0^{t_i} \sigma_j^2(s) \diff s \int_0^{t_k} \sigma_l^2(s) \diff s}}\,, \label{eqn:correlation_asian_basket}\end{equation}
for $(i,k) \in \{1,...,n\}^2$, $(j,l) \in\left\{1,...,m\right\}^2$. The Asian basket option is equivalent to a vanilla basket option on $\hat{S}_{i+n(j-1)}$ with weights $w_i \mu_j$.

\subsection{Strike}
The strike $K$ may be included in the spread sum. In the context of an Asian, it is equivalent to an asian observation at $t=0$ with weight $-K/S(0)$. 
In the context of a basket, it is equivalent to an additional asset of spot price $K$ with zero volatility and drift, weight $-1$, and correlation zero towards the other assets.

\section{Expansions}

\subsection{The proxy}
An early technique, first explored by \cite{vorst1992prices} to estimate the price of Asian options is to approximate the arithmetic average by the geometric average, because the geometric average of geometric Brownian motions is another geometric Brownian motion and can thus be priced with the usual Black-Scholes formula. This was extended to the more general cases of vanilla basket options by \cite{gentle1993basket}.

Under the risk-neutral measure $\mathbb{Q}$, the forward price for each asset follows
\begin{equation*}
	\diff F_i(t,T) = F_i(t,T) \sigma_i(t) \diff W_i(t)\,,
\end{equation*}
where $W_i$ is a $\mathbb{Q}$-Brownian motion, and we have $\diff W_i  \diff W_j = \rho_{i,j} \diff t$.
The payoff at maturity of the basket spread option where we allow for a multiplicative strike $\kappa \in \mathbb{R}$ reads
\begin{align*}
	V(T) &= \left[ \eta\left(\sum_{i=m_p}^m w_i S_i(T) + \kappa \sum_{i=1}^{m_p-1} w_i S_i(T)\right) \right]^+\\
	&= A_p \left[ \eta \left( \sum_{i=m_p}^m \tilde{a}_i S_i^\star(T) - {\kappa}^\star \sum_{i=1}^{m_p-1} |\tilde{a}_i| S_i^\star(T) \right) \right]^+	
\end{align*} with $\eta = \pm 1$ for a call (respectively a put) option and
\begin{align*}
	A_n &= -\sum_{i=1}^{m_p-1} w_i F_i(0,T)\,,\quad A_p = \sum_{i=m_p}^m w_i F_i(0,T)\,, \quad \kappa^\star = \frac{A_n}{A_p} \kappa\,,\\
	\tilde{a}_i &= \frac{w_i F_i(0,T)}{A_p(T)}\quad \textmd{ for } i \geq m_p \quad \textmd{ and } \quad \tilde{a}_i = \frac{w_i F_i(0,T)}{A_n(T)}\quad \textmd{ for } i < m_p\,,\\
	S_i^\star(t) &= \frac{S_i(t)}{F_i(0,t)} = e^{-\frac{1}{2}\int_{0}^{t} \sigma_i^2(s) \diff s + \int_{0}^{t} \sigma_i(s) \diff W_i(s)}\,.
\end{align*} 
In particular, we have $\tilde{a}_i < 0$ for $i < m_p$, and $A_n > 0$.
We approximate the arithmetic average by the geometric average leading to
\begin{equation}
	V(0) \approx  B(0,T) A \mathbb{E}\left[ \left( \eta\left(\prod_{i=m_p}^m  S_i^\star(T)^{a_i} - {\kappa}^\star\prod_{i=1}^{m_p-1}  S_i^\star(T)^{|a_i|}\right) \right)^+   \right]
\end{equation}
with $a_i = \tilde{a}_i$ as a specific case.

Let \begin{equation*}G(T,1,n) = \prod_{i=1}^n  S_i^\star(T)^{a_i}  = e^{ \sum_{i=1}^n -\frac{1}{2} a_i \int_{0}^{T} \sigma_i^2(s) \diff s + \int_{0}^{T} a_i \sigma_i(s) \diff W_i(s)}\end{equation*}
We have \begin{align}
	m(T,1,n) = \mathbb{E}\left[\ln G(T)\right] &=  -\frac{1}{2} \sum_{i=1}^n a_i v_i  \,,\label{eqn:ElogG}\\
	\tilde{\nu}^2(T,1,n) = \Var\left[\ln G(T)\right] &= \sum_{i,j=1}^n  a_i a_j \rho_{i,j}\int_0^T \sigma_i(s) \sigma_j(s) \diff s \label{eqn:VarG}\,.
	\end{align}
	with $v_i = \int_{0}^T \sigma_i^2(s) \diff s$. We will also define the total variances of the positive and negative parts 
	\begin{equation}
		\tilde\nu_n^2(T) = \tilde\nu(T,1,m_p-1)\,,\quad \tilde\nu_p^2 =  \tilde\nu(T,m_p,m)\,,\quad \tilde\nu^2(T)=\tilde\nu(T,1,m)\,.\label{eqn:nun_nup}
	\end{equation}
	
Instead of adjusting the strike to preserve the expectation of the term inside the max function as in \citep{vorst1992prices}, we adjust each geometric average by a factor (respectively $\alpha$ and $\beta$) to match the first moment of each corresponding arithmetic average:
\begin{equation}
	V(0) \approx  B(0,T) A_p \mathbb{E}\left[ \left(  \eta\left(\alpha\prod_{i=m_p}^m  S_i^\star(T)^{a_i} -  {\kappa}^\star\beta\prod_{i=1}^{m_p-1}  S_i^\star(T)^{|a_i|} \right)\right)^+ \right]\,,\label{eqn:vorst_proxy}
\end{equation}
with $\alpha = \frac{1}{\mathbb{E}[G(T,m_p,m)]}$, $\beta =  \mathbb{E}[G(T,1,m_p-1)]$ and $B(0,T)$ is the discount factor to the payment date. 

\cite{levy1992pricing} matches not only the mean but also the variance of the arithmetic average by a lognormal distribution. We have
\begin{align}
	\nu_{A_p}^2 &= \ln\left( \Var\left[A_p(T)\right] \right)= \ln\left( \sum_{i,j=m_p}^m a_i a_j e^{\rho_{i,j}\int_0^T \sigma_i(s) \sigma_j(s) \diff s}\right)\,,\\
	\nu_{A_n}^2 &= \ln\left( \Var\left[A_n(T)\right] \right)= \ln\left( \sum_{i,j=1}^{m_p-1} a_i a_j e^{\rho_{i,j}\int_0^T \sigma_i(s) \sigma_j(s) \diff s}\right)\,.
\end{align}
We may adjust the $a_i$ used in the geometric mean such that the variance of the geometric mean matches exactly the one of the arithmetic mean by using 
\begin{equation}
	a_i = \tilde{a}_i \frac{\nu_{A_p(T)}}{\tilde\nu_p(T)} \quad \textmd{ for } i \geq m_p\,, \quad \textmd{ and }\quad a_i = \tilde{a}_i \frac{\nu_{A_n}}{\tilde\nu_n(T)} \quad \textmd{ for } i < m_p\,.\label{eqn:ai_levy}
\end{equation}

In Identity \ref{identity:EGn}, we apply a change of measure to reduce Equation \ref{eqn:vorst_proxy} to a regular vanilla option pricing problem as in \citep{margrabe1978value}. This leads to
\begin{eqnarray}
	V(0) \approx  A_p  \cdot \Black\left(1, \kappa^\star, {\nu}^2, T\right)
\end{eqnarray}
where $\Black(F,K,v,T)$ denotes the Black-76 formula applied to the forward $F$, strike $K$ and total variance $v$ to maturity $T$ (Equation \ref{eqn:black}) and $\nu = \tilde{\nu}$ for the geometric approximation, $\nu$ computed from $a_i$ for the Levy approximation.

We will adopt the Vorst geometric formula with adjusted forward as proxy for our stochastic Taylor expansion. Instead of our normalization factors $\alpha$ and $\beta$, \cite{krekel2003pricing} employs an additive shift to adjust the expectation of the whole spread, as does \cite{vorst1992prices} for regular Asian options. It would lead to slightly more complicated equations and exhibit a worse accuracy.

In the context of standard Asian options (no spread), the Vorst Levy formula corresponds exactly to the Levy approximation, but defines a stochastic process based on the individual underlying asset processes. This will allow to define  stochastic expansions based on this alternative proxy as well.

Both approximations have a major advantage compared to a shifted lognormal moment matching approach, such as in \citep{borovkova2007closed}: each sum of lognormal distributions is approximated by a distribution defined on the positive real line. In contrast, for example in the case of a basket with positive weights, the shifted lognormal moment matching may lead to a positive shift, which will allow for negative values of the basket price, which is not realistic. With weights of mixed signs, the shifted lognormal approximation requires special handling to capture negative skews, as described in \citep{borovkova2007closed}.

\subsection{First order expansion}
Let $h(x) = \max(\eta x, 0)$, we apply a Taylor expansion of order-1 on  $h$:
\begin{multline}
	\mathbb{E}\left[ h\left(\sum_{i=m_p}^m \tilde{a}_i S_{i}^{\star}(T) -\kappa^\star \sum_{i=1}^{m_p-1} |\tilde{a}_i| S_{i}^{\star}(T)\right) \right] = \mathbb{E}\left[ h\left(G_p^\star(T) - \kappa^\star G_n^\star(T) \right)\right] \\
	+ 	\mathbb{E}\left[ \left( \sum_{i=m_p}^m \tilde{a}_i S_{i}^{\star}(T) -\kappa^\star \sum_{i=1}^{m_p-1} |\tilde{a}_i| S_{i}^{\star}(T) -G_p^\star(T) + \kappa^\star G_n^\star(T)\right) h'\left(G_p^\star(T) - \kappa^\star G_n^\star(T)\right)\right]\\ + \Error_2(h)
\end{multline}
where $\Error_2(h) \leq_c 2 \left(||  \sum_{i=m_p}^m \tilde{a}_i S_{i}^{\star}(T) -\kappa^\star \sum_{i=1}^{m_p-1} |\tilde{a}_i| S_{i}^{\star}(T) -G_p^\star(T) + \kappa^\star G_n^\star(T) ||_3\right)^2$. 

Let $\tilde{a}_i^\star = \tilde{a}_i$ for $i \geq m_p$ and $\tilde{a}_i^\star = \kappa^\star \tilde{a}_i$ for $i < m_p$, we may rewrite the second term using a derivative with regards to $\kappa^\star$, and interchange derivation and expectation (see \cite{etore2012stochastic})  to obtain
\begin{multline}	\mathbb{E}\left[ \left( \sum_{i=1}^m \tilde{a}_i^\star S_{i}^{\star}(T) - G_p^\star(T) + \kappa^\star G_n^\star(T)\right) h'\left(G_p^\star(T) - \kappa^\star G_n^\star(T)\right)\right]\\  = -\frac{\partial}{\partial \kappa^\star} \mathbb{E}\left[ \frac{ \sum_{i=1}^m \tilde{a}_i^\star S_{i}^{\star}(T) -  G_p^\star(T) + \kappa^\star G_n^\star(T)}{ G_n^\star(T)} h\left( G_p^\star(T) - \kappa^\star G_n^\star(T)\right)\right] \label{eqn:first_order_correction_raw}
	\end{multline}

We use Identities \ref{identity:EGn}, \ref{identity:EGp} and \ref{identity:ESi} to obtain the first order expansion
\begin{multline}
	V_\textmd{VG1} = A_p \cdot \Black(1,\kappa^\star,\nu^2,T) \\
	+ A_p\frac{\partial}{\partial \kappa^\star}\Black\left(e^{\nu^2},\kappa^\star,\nu^2,T\right)
	- \kappa^\star A_p\frac{\partial}{\partial \kappa^\star}\Black\left(1,\kappa^\star,\nu^2,T\right)\\
	 - A_p\sum_{i=1}^n \tilde{a}_i^\star \frac{\partial}{\partial \kappa^\star}\Black\left(e^{-\frac{\nu_p^2(T)-\nu_n^2(T)-\nu^2(T)}{2}+\bar{v}_i},\kappa^\star,\nu^2,T\right) \,.
\end{multline}
with $\bar{v}_i$ defined in Equation \ref{eqn:vibar_vil}.
The relevant derivatives of the Black-76 formula are given in Appendix \ref{sec:appendix_black_der}. The formula with the Vorst Levy proxy is exactly the same, but uses different $\nu$ and $a_i$.

Interestingly, while our derivation is different, the first order expansion on the geometric average "VG1" is the same as the extended Bjersksund Stensland approximation of \cite{pellegrino2016general}.

\subsection{Second order expansion}
The second-order Taylor expansion involves the additional term
\begin{equation*}
	\frac{1}{2}\frac{\partial^2}{\partial {\kappa^\star}^2} \mathbb{E}\left[ \frac{\left( \sum_{i=1}^m \tilde{a}_i^\star S_{i}^{\star}(T) -  G_p^\star(T) + \kappa^\star G_n^\star(T)\right)^2}{ {G_n^\star}(T)} h\left( \frac{G_p^\star(T)}{G_n^\star(T)} - \kappa^\star\right)\right]
\end{equation*}
We decompose it into six terms:
\begin{multline*}
	\frac{\left( \sum_{i=1}^m \tilde{a}_i^\star S_{i}^{\star}(T) -  G_p^\star(T) + \kappa^\star G_n^\star(T)\right)^2}{ {G_n^\star}(T)} =  G_p^\star(T)\frac{G_p^\star(T)}{G_n^\star(T)} + {\kappa^\star}^2 G_n^\star(T)\\ - 2\kappa^\star G_p^\star (T) - 2 \left( \sum_{i=1}^m\tilde{a}_i^\star S_{i}^{\star}(T) \right)\frac{G_p^\star}{G_n^\star} + 2 \kappa^\star\left( \sum_{i=1}^m\tilde{a}_i^\star S_{i}^{\star}(T)\right) + \sum_{i=1}^m \sum_{j=1}^m \tilde{a}_i^\star \tilde{a}_j^\star \frac{S_{i}^{\star}(T)S_{J}^{\star}(T)}{G_n^\star(T)}\,.
\end{multline*}

We then use Identities \ref{identity:EGn} to \ref{identity:ESi2overGn} to find
our formula for the second order expansion:
\begin{multline}
	V_\textmd{VG2} = V_\textmd{VG1}
	 +\frac{ A_p e^{\nu^2(T)}}{2}\frac{\partial^2}{\partial {\kappa^{\star}}^2}\Black\left(e^{2\nu^2}, \kappa^\star, \nu^2, T\right)
	 	 +\frac{\kappa^2 A_p}{2}\frac{\partial^2}{\partial {\kappa^{\star}}^2} \Black\left(1, \kappa^\star, {\nu}^2, T\right)\\
	 -\kappa^\star A_p\frac{\partial^2}{\partial {\kappa^{\star}}^2}\Black\left(e^{\nu^2}, \kappa^\star, \nu^2, T\right)
	 -{A_p} \sum_{i=1}^m \tilde{a}_i^\star e^{\nu_n^2-\nu_{pn}^2+\bar{v}_i} \frac{\partial^2}{\partial {\kappa^\star}^2} \Black\left(e^{\nu^2+\nu_n^2-\nu_{pn}^2+\bar{v}_i}, \kappa^\star, \nu^2, T\right)\\
	+\kappa^\star{A_p} \sum_{i=1}^m \tilde{a}_i^\star  \frac{\partial^2}{\partial {K^\star}^2} \Black\left(e^{-\frac{\nu_p^2-\nu_n^2-\nu^2}{2}+\bar{v}_i},\kappa^\star,\nu^2,T\right)\\
	+\frac{A_p}{2} \sum_{i,j=1}^n \tilde{a}_i^\star \tilde{a}_j^\star  e^{\nu_n^2+{v}_{i,j}+\bar{v}_{i}^- + \bar{v}_{j}^-(T)}  \frac{\partial^2}{\partial {\kappa^\star}^2} \Black\left(e^{\bar{v}_{i}+\bar{v}_j -2\nu_{np}^2 +2\nu_n^2} , \kappa^\star, \nu^2, T\right)\,,
\end{multline}
with $\bar{v}_i$, $v_{i,j}$ defined in Equation \ref{eqn:vibar_vil} and $\bar{v}_i^-$ defined in Equation \ref{eqn:vin}. 
\subsection{Third order expansion}
The third order expansion adds the term
\begin{align*}
-	\frac{1}{6}\frac{\partial^3}{\partial {\kappa^\star}^3} \mathbb{E}\left[ \frac{\left(\sum_{i=1}^m \tilde{a}_i^\star S_{i}^{\star}(T) -  G_p^\star(T) + \kappa^\star G_n^\star(T)\right)^3}{{G_n^\star(T)}^2} h\left(\alpha \frac{G_p^\star(T)}{G_n^\star(T)} - \kappa^\star\right)\right]
\end{align*}
We decompose it into the following terms:
\begin{multline*}
	\frac{\left(\sum_{i=1}^m \tilde{a}_i^\star S_{i}^{\star}(T) -  G_p^\star(T) + \kappa^\star G_n^\star(T)\right)^3}{{G_n^\star(T)}^2} = 
	- \frac{{G_p^\star}^3}{{G_n^\star}^2}
	+ 3\kappa^\star\frac{{G_p^\star}^2}{{G_n^\star}}
	- 3{\kappa^\star}^2 G_p^\star
	+  {\kappa^\star}^3 {G_n^\star}\\
	+ 3\sum_{i=1}^m \tilde{a}_i^\star  \frac{S_{i}^{\star}{G_p^\star}^2(T)}{{G_n^\star}^2}
	+3{\kappa^\star}^2\sum_{i=1}^m \tilde{a}_i^\star  S_{i}^{\star} - 6\kappa^\star\sum_{i=1}^m \tilde{a}_i^\star  \frac{S_{i}^{\star}{G_p^\star}(T)}{{G_n^\star}}\\
	 - 3\sum_{i,j=1}^m \tilde{a}_i^\star \tilde{a}_j^\star  \frac{S_{i}^{\star}S_{j}^{\star}G_p^\star(T)}{{G_n^\star}^2}
	 +3\kappa^\star\sum_{i,j=1}^m \tilde{a}_i^\star \tilde{a}_j^\star  \frac{S_{i}^{\star}S_{j}^{\star}}{{G_n^\star}}
	+ \sum_{i,j,k=1}^m \tilde{a}_i^\star \tilde{a}_j^\star \tilde{a}_k^\star \frac{S_{i}^{\star}S_{j}^{\star}S_{k}^{\star}(T)}{{G_n^\star}^2}
\end{multline*}
We use Identities \ref{identity:EGn} to \ref{identiy:SiSjSkoverGn2} to obtain the third-order formula
\newtcolorbox{mymathbox}[1][]{colback=white, sharp corners, #1}

\begin{multline}
	V_\textmd{VG3} = V_\textmd{VG2}\\
	+\frac{A_p e^{3\nu^2(T)}}{6} \frac{\partial^3 \Black}{\partial {\kappa^{\star}}^3}\left(e^{3\nu^2(T)}, \kappa^\star, \nu^2, T\right)
	-\frac{A_p \kappa^\star e^{\nu^2}}{2} \frac{\partial^3 \Black}{\partial {\kappa^{\star}}^3} \left(e^{2\nu^2}, \kappa^\star, \nu^2, T\right)\\
	+\frac{A_p {\kappa^\star}^2}{2} \frac{\partial^3 \Black}{\partial {\kappa^{\star}}^3} \left(e^{\nu^2}, \kappa^\star, \nu^2, T\right)
	-\frac{A_p {\kappa^\star}^3}{6} \frac{\partial^3 \Black}{\partial {\kappa^{\star}}^3}\left(1, \kappa^\star, {\nu}^2, T\right)\\
	-\frac{A_p}{2}\sum_{i=1}^m \tilde{a}_i^\star e^{\nu^2+2\nu_n^2-2\nu_{pn}^2+2\bar{v}_i}\frac{\partial^3 \Black}{\partial {\kappa^{\star}}^3}\left(e^{2\nu^2+\nu_n^2-\nu_{pn}^2+\bar{v}_i}, \kappa^\star, \nu^2, T\right)\\
	-\frac{A_p {\kappa^\star}^2}{2}\sum_{i=1}^m \tilde{a}_i^\star \frac{\partial^3 \Black}{\partial {\kappa^{\star}}^3}\left(e^{-\frac{\nu_p^2-\nu_n^2-\nu^2}{2}+\bar{v}_i},\kappa^\star,\nu^2,T\right) \\
	+A_p {\kappa^\star}\sum_{i=1}^m \tilde{a}_i^\star e^{\nu_n^2-\nu_{pn}^2+\bar{v}_i}\frac{\partial^3 \Black}{\partial {\kappa^{\star}}^3}\left(e^{\nu^2+\nu_n^2-\nu_{pn}^2+\bar{v}_i}, \kappa^\star, \nu^2, T\right)\\
	+\frac{A_p}{2}\sum_{i,j=1}^m \tilde{a}_i^\star \tilde{a}_j^\star  e^{3\nu_n^2-2\nu_{pn}^2+\bar{v}_i+\bar{v}_j+{v}_{i,j}+\bar{v}_{i}^- + \bar{v}_{j}^-(T)} \frac{\partial^3 \Black}{\partial {\kappa^\star}^3} \left(e^{\bar{v}_{i}+\bar{v}_j +\nu^2 -2\nu_{pn}^2 +2\nu_n^2} , \kappa^\star, \nu^2, T\right)\\
	-\frac{A_p \kappa^\star}{2}\sum_{i,j=1}^m \tilde{a}_i^\star \tilde{a}_j^\star  e^{\nu_n^2+{v}_{i,j}+\bar{v}_{i}^- + \bar{v}_{j}^-(T)}\frac{\partial^3 \Black}{\partial {\kappa^\star}^3}\left(e^{\bar{v}_{i}+\bar{v}_j -2\nu_{np}^2 +2\nu_n^2} , \kappa^\star, \nu^2, T\right)\\
	-\frac{A_p}{6}  \sum_{i,j,k=1}^m \tilde{a}_i^\star \tilde{a}_j^\star \tilde{a}_k^\star	e^{{v}_{i,j}+{v}_{k,i}+{v}_{k,j}+2\bar{v}_{i}^- + 2\bar{v}_{j}^- +2\bar{v}_{k}^- + 3\nu_n^2} \frac{\partial^3 \Black}{\partial {\kappa^\star}^3}\left(e^{\bar{v}_{i}+\bar{v}_j+\bar{v}_{k} -3\nu_{np}^2 +3\nu_n^2} , \kappa^\star, \nu^2, T\right)\nonumber\,,
\end{multline}
with $\bar{v}_i$, $v_{i,j}$ defined in Equation \ref{eqn:vibar_vil} and $\bar{v}_i^-$ defined in Equation \ref{eqn:vin}.
\subsection{Symmetries}
We may exploit the symmetries in the double and triple sums to reduce the number terms being summed and speed up the corresponding algorithm.
In particular,
\begin{multline*}
 \sum_{i,j=1}^n  \tilde{a}_i \tilde{a}_j  e^{v_{i,j}}  \frac{\partial^2}{\partial {K^\star}^2}\Black\left(e^{\bar{v}_i + \bar{v}_j},K^\star,\nu^2,T\right) = \sum_{i=1}^n  \tilde{a}_i^2  e^{v_{i,i}}  \frac{\partial^2}{\partial {K^\star}^2}\Black\left(e^{2\bar{v}_i },K^\star,\nu^2,T\right)\\ + 2\sum_{i=1}^n \sum_{j=1}^{i-1}   \tilde{a}_i \tilde{a}_j  e^{v_{i,j}}  \frac{\partial^2}{\partial {K^\star}^2}\Black\left(e^{\bar{v}_i + \bar{v}_j},K^\star,\nu^2,T\right)\,.
\end{multline*}
and
\begin{multline*}
 \sum_{i,j,l=1}^n \tilde{a}_i \tilde{a}_j \tilde{a}_l e^{v_{i,l}+v_{j,l}+v_{i,j}} 	 \frac{\partial^3}{\partial {K^\star}^3}\Black\left(e^{\bar{v}_i+\bar{v}_j+\bar{v}_l},K^\star,\nu^2,T\right) =\\
6 \sum_{i=1}^{n} \sum_{j=1}^{i-1} \sum_{l=1}^{j-1} \tilde{a}_i \tilde{a}_j \tilde{a}_l e^{v_{i,l}+v_{j,l}+v_{i,j}} 	 \frac{\partial^3}{\partial {K^\star}^3}\Black\left(e^{\bar{v}_i+\bar{v}_j+\bar{v}_l},K^\star,\nu^2,T\right)\\
+ 3 \sum_{i=1}^n \sum_{j=1}^{i-1} \tilde{a}_i \tilde{a}_j^2  e^{2v_{i,j}+v_{j,j}} 	 \frac{\partial^3}{\partial {K^\star}^3}\Black\left(e^{\bar{v}_i+2\bar{v}_j},K^\star,\nu^2,T\right)\\
+ 3 \sum_{i=1}^n \sum_{l=1}^{i-1} \tilde{a}_i^2 \tilde{a}_l e^{2v_{i,l}+v_{i,i}} 	 \frac{\partial^3}{\partial {K^\star}^3}\Black\left(e^{2\bar{v}_i+\bar{v}_l},K^\star,\nu^2,T\right)\\
+ \sum_{i=1}^n \tilde{a}_i^3 e^{3v_{i,i}} 	 \frac{\partial^3}{\partial {K^\star}^3}\Black\left(e^{3\bar{v}_i},K^\star,\nu^2,T\right)\,.
\end{multline*}

\section{Numerical examples}
For regular Asian and basket options (i.e., with positive weights only), the approximations are equivalent to the ones of \cite{lefloc2024stochastic}. We will thus consider only cases of spread options, (i.e., with some negative weights).

\subsection{Spread options}
We start with the example presented in \cite[Table 1]{deelstra2010pricing}. It consists in pricing a call spread option of maturity 1 year paying $(S_2-S_1-K)^+$, using a risk-free interest rate $r=5\%$ on two assets $S_1$ and $S_2$ with $S_1(0) = 200$, $S_2(0)=100$ and $\sigma_1=60\%, \sigma_2=60\%$ with correlation $\rho_{1,2} = 28\%$ for a range of strikes $K$. Our Monte-Carlo prices differs slightly from the ones presented in \cite{deelstra2010pricing} as we employed a Monte-Carlo simulation with a digitally scrambled Sobol sequence \citep{l2002recent} on 64 millions paths. Compared to the Monte-Carlo simulation of \cite{deelstra2010pricing} using 100 millions of paths, our prices have a better accuracy, which is important since they are used to compute the root mean square errors (RMSE) and maximum absolute errors (MAE) of the various approximations.
On this example, the basket skewness $\mu_3$ is negative, where
\begin{equation*}
	\mu_3  = \frac{\mathbb{E}\left[\left(\sum_{i=1}^m w_i S_i(T)- \mathbb{E}\left[\sum_{i=1}^m w_i S_i(T)\right]\right)^3\right]}{\left(\mathbb{E}\left[\left(\sum_{i=1}^m w_i S_i(T)\right)^2\right] - \mathbb{E}\left[\sum_{i=1}^m w_i S_i(T)\right]^2 \right)^{\frac{3}{2}}  }\,.
\end{equation*}

As per Table \ref{tbl:deelstra1}, the proxy prices (VG0) are already quite accurate, much better than the shifted lognormal (SLN) approximation of \cite{borovkova2007closed}. The first order (VG1) does not improve the accuracy, while the second order (VG2) is an order of magnitude more accurate. The third order (VG3) allows to gain another order of magnitude in accuracy. The most accurate is the improved comonotonic upper bound (ICUB) approximation of \cite{deelstra2010pricing}.
\begin{table}[H]
	\caption{Call spread option of maturity 1 year paying $(S_2-S_1-K)^+$, on two assets $S_1$ and $S_2$ with $S_1(0) = 200$, $S_2(0)=100$ and $\sigma_1=60\%, \sigma_2=60\%$ with correlation $\rho_{1,2} = 28\%$. This example has negative basket skewness\label{tbl:deelstra1}}
	\centering{
		\begin{tabular}{@{}l r r r r r r  r@{}}\toprule
			$K$ &  VG0 & VG1 & VG2 & VG3 & SLN & ICUB & MC \\ \midrule
			-70 & 28.9222& 28.9251& 29.0906& 29.0853& 31.0619& 29.0854& 29.0854\\
-80 & 33.4520& 33.4313& 33.6233& 33.6148& 35.8096& 33.6150& 33.6150\\
-90 & 38.3631& 38.3258& 38.5397& 38.5276& 40.8772& 38.5281& 38.5281\\
-100 & 43.6358& 43.5882& 43.8192& 43.8031& 46.2500& 43.8043& 43.8043\\
-110 & 49.2485& 49.1961& 49.4394& 49.4194& 51.9127& 49.4212& 49.4212\\
-120 & 55.1792& 55.1263& 55.3770& 55.3535& 57.8497& 55.3561& 55.3561\\
-130 & 61.4056& 61.3551& 61.6093& 61.5826& 64.0453& 61.5861& 61.5861\\
			\midrule
			RMSE &  0.1701 & 0.2084 & 0.0158 & 0.0018 & 2.3512 & 0.0000 & 0.0000 			\\
			MAE &  0.1805 & 0.2310 & 0.0232 & 0.0034 & 2.4936 & 0.0000 & 0.0000\\
			\bottomrule
	\end{tabular}}
\end{table}

The second example originally from \cite[Table 2]{deelstra2010pricing} is a similar spread option but using $S_1(0)=40, S_2(0)=100$, $\sigma_1=17\%, \sigma_2=40\%$ and $\rho_{1,2}=12\%$. The smaller vols results in a much more accurate Taylor expansion, and the second order expansion is the most accurate on this example (Table \ref{tbl:deelstra2}). The ICUB approximation is less accurate as in the context of a negative basket skewness.
\begin{table}[H]
	\caption{Call spread option of maturity 1 year paying $(S_2-S_1-K)^+$, on two assets $S_1$ and $S_2$ with $S_1(0)=40, S_2(0)=100$, $\sigma_1=17\%, \sigma_2=40\%$ and $\rho_{1,2}=12\%$. This example has positive basket skewness.\label{tbl:deelstra2}}
	\centering{
		\begin{tabular}{@{}l r r r r r r r@{}}\toprule
			$K$ &  VG0 & VG1 & VG2 & VG3 & SLN & ICUB & MC \\ \midrule
			45 & 24.5969& 24.5971& 24.5982& 24.5982& 24.6096& 24.5975& 24.5982\\
			50 & 21.8235& 21.8236& 21.8247& 21.8247& 21.8441& 21.8240& 21.8247\\
			55 & 19.3074& 19.3074& 19.3086& 19.3086& 19.3342& 19.3079& 19.3086\\
			60 & 17.0380& 17.0380& 17.0391& 17.0391& 17.0692& 17.0384& 17.0391\\
			65 & 15.0018& 15.0017& 15.0029& 15.0029& 15.0355& 15.0022& 15.0029\\
			70 & 13.1831& 13.1830& 13.1842& 13.1842& 13.2178& 13.1835& 13.1842\\
			75 & 11.5653& 11.5652& 11.5663& 11.5663& 11.5997& 11.5656& 11.5663\\
							\midrule
			RMSE &0.0011 & 0.0011 & 0.0000 & 0.0000 & 0.0277 & 0.0007 & 0.0000 \\
			MAE & 0.0013 & 0.0012 & 0.0000 & 0.0000 & 0.0336 & 0.0007 & 0.0000 \\
			\bottomrule
	\end{tabular}}
\end{table}

\subsection{Basket spread options}
We now consider the example from \cite[Table 4]{deelstra2010pricing} of a basket spread option with payoff $(S_3-S_2-S_1-K)^+$ at maturity $T=1$, where $S_3(0)=100$, $S_2(0)=24$, $S_1(0)=46$, $\sigma_3=40\%$, $\sigma_2=22\%$, $\sigma_1=30\%$, $\rho_{3,2}=17\%$, $\rho_{1,3}=91\%$, $\rho_{2,1}=41\%$. On this first example (Table \ref{tbl:deelstra4}), the second-order and third-order reach beyond our Monte-Carlo reference prices accuracy. 
\begin{table}[H]
	\caption{Basket spread option with payoff $(S_3-S_2-S_1-K)^+$ at maturity $T=1$, where $S_3(0)=100$, $S_2(0)=24$, $S_1(0)=46$, $\sigma_3=40\%$, $\sigma_2=22\%$, $\sigma_1=30\%$, $\rho_{3,2}=17\%$, $\rho_{1,3}=91\%$, $\rho_{2,1}=41\%$.\label{tbl:deelstra4}}
	\centering{
		\begin{tabular}{@{}l r r r r r r r@{}}\toprule
			$K$ &  VG0 & VG1 & VG2 & VG3 & SLN & HICUB & MC \\ \midrule
		25.0 & 20.7593& 20.7611& 20.7646& 20.7646& 20.8073& 20.7637& 20.7645\\
30.0 & 17.6876& 17.6884& 17.6931& 17.6931& 17.7711& 17.6921& 17.6931\\
35.0 & 14.9531& 14.9532& 14.9590& 14.9589& 15.0630& 14.9580& 14.9591\\
40.0 & 12.5514& 12.5512& 12.5577& 12.5576& 12.6769& 12.5566& 12.5579\\
45.0 & 10.4677& 10.4676& 10.4746& 10.4744& 10.5983& 10.4734& 10.4747\\
50.0 & 8.6800& 8.6799& 8.6871& 8.6870& 8.8065& 8.6859& 8.6873\\
55.0 & 7.1612& 7.1613& 7.1684& 7.1682& 7.2767& 7.1671& 7.1685\\
		\midrule
			RMSE &0.0829 & 0.0054 & 0.0001 & 0.0000 & 0.0455 & 0.0959 & 0.0000 \\
			MAE &0.1351 & 0.0059 & 0.0002 & 0.0000 & 0.0566 & 0.1625 & 0.0000 \\
			\bottomrule
	\end{tabular}}
\end{table}

Next, we consider a similar example \citep[Table 7]{deelstra2010pricing}, presented in Table \ref{tbl:deelstra7} with higher volatilities:  $S_3(0)=100$, $S_2(0)=63$, $S_1(0)=12$, $\sigma_3=21\%$, $\sigma_2=34\%$, $\sigma_1=63\%$, $\rho_{3,2}=87\%$, $\rho_{1,3}=30\%$, $\rho_{2,1}=43\%$.
\begin{table}[H]
	\caption{Basket spread option with payoff $(S_3-S_2-S_1-K)^+$ at maturity $T=1$, where $S_3(0)=100$, $S_2(0)=63$, $S_1(0)=12$, $\sigma_3=21\%$, $\sigma_2=34\%$, $\sigma_1=63\%$, $\rho_{3,2}=87\%$, $\rho_{1,3}=30\%$, $\rho_{2,1}=43\%$.\label{tbl:deelstra7}}
	\centering{
		\begin{tabular}{@{}l r r r r r r r@{}}\toprule
			$K$ &  VG0 & VG1 & VG2 & VG3 & SLN & HICUB & MC \\ \midrule
			2.5 & 23.4535& 23.5493& 23.5998& 23.5941& 23.1681& 23.5138& 23.5938\\
10.0 & 17.0131& 17.1596& 17.2144& 17.2040& 16.8591& 17.1373& 17.2063\\
17.5 & 11.2434& 11.3588& 11.4175& 11.4084& 11.3394& 11.3873& 11.4112\\
25.0 & 6.5548& 6.5418& 6.6041& 6.6006& 6.9203& 6.6584& 6.6023\\
32.5 & 3.2670& 3.1203& 3.1849& 3.1870& 3.7629& 3.3147& 3.1877\\
40.0 & 1.3636& 1.1852& 1.2504& 1.2527& 1.7925& 1.3853& 1.2518\\
47.5 & 0.4758& 0.3537& 0.4072& 0.4032& 0.7369& 0.4913& 0.4024\\
			\midrule
			RMSE &  0.1263 & 0.0560 & 0.0050 & 0.0016 & 0.4041 & 0.0900 & 0.0000 \\
			MAE &0.1932 & 0.0675 & 0.0081 & 0.0028 & 0.5752 & 0.1335 & 0.0000\\
			\bottomrule
	\end{tabular}}
\end{table}
The second order expansion is more accurate than the hybrid ICUB by a factor of 10. The latter was found to be the most accurate in \citep{deelstra2010pricing}. The third order expansion further improves accuracy by a factor four.

\subsection{Asian basket spread options}
The option’s maturity is set equal to one year and the averaging is taken over the last 30 days, which is a common practice in energy markets.

The first example comes from \cite[Table 8]{deelstra2010pricing} and consists in an Asian basket spread option with payoff $\left(\sum_{i=1}^n w_i S_2(t_i)-\sum_{i=1}^n w_i S_1(t_i) - K\right)^+$ and $S_2(T)=100$, $S_1(T)=60$, $\sigma_2 = 33\%$, $\sigma_1= 25\%$, $\rho_{1,2}=40\%$, $w_i=1/30$, $t_i=T-(i-1)/30$, $n=30$.
In order to assess the true accuracy of the third-order expansion, we recalculated the reference Monte-Carlo prices with a higher accuracy than \cite{deelstra2010pricing}, by using Brownian-Bridge variance reduction on a randomized Sobol sequence: the third order expansion reaches the limits of our quasi Monte-Carlo simulation accuracy with more than 32 millions paths (\ref{tbl:deelstra8}).
\begin{table}[H]
	\caption{Asian basket spread option with payoff $\left(\sum_{i=1}^n w_i S_2(t_i)-\sum_{i=1}^n w_i S_1(t_i) - K\right)^+$ with $S_2(T)=100$, $S_1(T)=60$, $\sigma_2 = 33\%$, $\sigma_1= 25\%$, $\rho_{1,2}=40\%$.\label{tbl:deelstra8}}
	\centering{
		\begin{tabular}{@{}l r r r r r r r@{}}\toprule
			$K$ &  VG0 & VG1 & VG2 & VG3 & SLN & HICUB & MC \\ \midrule
			25.0 & 20.7593& 20.7611& 20.7646& 20.7646& 20.8073& 20.7637& 20.7646\\
30.0 & 17.6876& 17.6884& 17.6931& 17.6931& 17.7711& 17.6921& 17.6931\\
35.0 & 14.9531& 14.9532& 14.9590& 14.9589& 15.0630& 14.9580& 14.9589\\
40.0 & 12.5514& 12.5512& 12.5577& 12.5576& 12.6769& 12.5566& 12.5577\\
45.0 & 10.4677& 10.4676& 10.4746& 10.4744& 10.5983& 10.4734& 10.4745\\
50.0 & 8.6800& 8.6799& 8.6871& 8.6870& 8.8065& 8.6859& 8.6870\\
55.0 & 7.1612& 7.1613& 7.1684& 7.1682& 7.2767& 7.1671& 7.1683\\
						\midrule
			RMSE & 0.0063 & 0.0060 & 0.0001 & 0.0000 & 0.1031 & 0.0010 & 0.0000 			\\
			MAE &0.0070 & 0.0070 & 0.0002 & 0.0000 & 0.1238 & 0.0012 & 0.0000 \\
			\bottomrule
	\end{tabular}}
\end{table}

The next example corresponds to \cite[Table 10]{deelstra2010pricing} and is an Asian basket spread option with payoff \begin{equation*}V(T)=\left(\sum_{i=1}^n w_i S_3(t_i)-\sum_{i=1}^n w_i S_2(t_i)-\sum_{i=1}^n w_i S_1(t_i) - K\right)^+\,,\end{equation*} with $S_3(0)=100$, $S_2(0)= 50$, $S_1(0) = 25$, $\sigma_3 = 35$, $\sigma_2=30\%$, $\sigma_1=25\%$, $\rho_{2,3} = 30\%$, $\rho_{1,3} = 80\%$,   $\rho_{1,2} = 70\%$.

On this second example, the proxy is already as accurate as the hybrid ICUB approximation, and the second order expansion reaches the limits of the Monte-Carlo simulation with 300 millions of paths from \cite{deelstra2010pricing}, while the third-oder reaches the limits of our quasi Monte-Carlo simulation (Table \ref{tbl:deelstra10}). This confirms the findings of \cite{lefloc2024stochastic}, which noticed that stochastic Taylor expansions were particularly accurate for Asian options, as the averaging reduces the effective volatility.

\begin{table}[H]
	\caption{Asian basket spread option with payoff $\left(\sum_{i=1}^n w_i (S_3(t_i) - S_2(t_i)- S_1(t_i)) - K\right)^+$.\label{tbl:deelstra10}}
	\centering{
		\begin{tabular}{@{}l r r r r r r r@{}}\toprule
			$K$ &  VG0 & VG1 & VG2 & VG3 & SLN & HICUB & MC \\ \midrule
			10.0 & 20.5577& 20.5991& 20.6010& 20.6010& 20.7157& 20.5521& 20.6010\\
			15.0 & 17.5270& 17.5514& 17.5546& 17.5545& 17.7346& 17.5209& 17.5546\\
			20.0 & 14.8298& 14.8364& 14.8409& 14.8409& 15.0677& 14.8236& 14.8409\\
			25.0 & 12.4617& 12.4516& 12.4575& 12.4574& 12.7097& 12.4559& 12.4574\\
			30.0 & 10.4079& 10.3836& 10.3906& 10.3906& 10.6478& 10.4030& 10.3906\\
			35.0 & 8.6463& 8.6108& 8.6187& 8.6186& 8.8635& 8.6425& 8.6186\\
			40.0 & 7.1499& 7.1063& 7.1147& 7.1145& 7.3342& 7.1472& 7.1145\\
			
			\midrule
			RMSE &0.0270 & 0.0059 & 0.0001 & 0.0000 & 0.2188 & 0.0283 & 0.0000			\\
			MAE &0.0434 & 0.0082 & 0.0002 & 0.0000 & 0.2572 & 0.0489 & 0.0000\\
			\bottomrule
	\end{tabular}}
\end{table}

\subsection{Asian spread options}
We consider the example from \cite{krekel2003pricing} of equity Asian spread options with payoff 
$\left(\sum_{i=27}^{52} w_i S(t_i) - K\sum_{j=1}^{26} w_j S(t_j)\right)^+$ where $w_i=1/26$, $t_i=\frac{i}{52}$ for $i \leq 26$ and $t_i = 4.5 + \frac{i-26}{52}$ for $i>26$, spot price $S(0)=100$, interest rate $r=5\%$, dividend yield $q=1\%$, for a range of multiplicative strikes $K$. We consider the most challenging case, that is the seasoned option (at $t=0.25$) with high volatility $\sigma=70\%$. Thus, thirteen observations are known, and their fixings are set to the original forward prices $F(0,t_j)$. In this example, the forward prices are all set from $t=0$ and not from $t=0.25$. We compute reference Monte-Carlo prices using Brownian-bridge variance reduction on a digitally scrambled Sobol sequence of 1 billion paths.

The proxy is already more accurate than their moment matching formula\footnote{Their moment matching formula consists in a one-dimensional integration of two lognormal moment matched distributions, one for the averaging spot part, one for the averaging strike part.} "MM" , which is the most accurate method presented in \citep[Table 2]{krekel2003pricing}. The second order expansion is around 40 times more accurate, and the third order reaches the limit of our Monte-Carlo simulation accuracy, with a maximum error below $0.002$  basis point of the spot price (Table \ref{tbl:krekel2}).
\begin{table}[H]
	\caption{Prices of seasoned Asian spread options with payoff $\left(\sum_{i=27}^{52} w_i S(t_i) - K\sum_{j=1}^{26} w_j S(t_j)\right)^+$ for a range of multiplicative strikes $K$.\label{tbl:krekel2}}
	\centering{
		\begin{tabular}{@{}l r r r r r r@{}}\toprule
			$K$ &  VG0 & VG1 & VG2 & VG3 & MM & QMC \\ \midrule
0.5 & 68.69262& 68.68561& 68.68818& 68.68812& 68.69909& 68.68810\\
0.8 & 59.55470& 59.54542& 59.54909& 59.54901& 59.56331& 59.54899\\
1.0 & 54.79849& 54.78831& 54.79251& 54.79242& 54.80797& 54.79240\\
1.2 & 50.77850& 50.76772& 50.77233& 50.77224& 50.78856& 50.77223\\
1.5 & 45.76550& 45.75420& 45.75928& 45.75919& 45.77606& 45.75917\\

\midrule
RMSE & 0.00582 & 0.00402 & 0.00010 & 0.00002 & 0.01497 & 0.00000 \\
MAE & 0.00633 & 0.00497 & 0.00011 & 0.00002 & 0.01689 & 0.00000 \\
		\bottomrule
	\end{tabular}}
\end{table}
Those observations also holds true for the newly issued Asian spread option case.
\section{Conclusion}
The second-order expansion is found to be the most accurate, by far, compared to previously published approximations. The third-order expansion goes further beyond, and reaches the limits of our reference quasi Monte-Carlo simulation accuracy for a much lower computational cost. This shows the power of stochastic Taylor expansions: they are very simple and yet lead to surprisingly accurate approximations. Their range of applicability is also great. So far, we have covered discrete cash dividends, Asian options, basket options (to some extent) and Asian basket spread options under the Black-Scholes model with a term-structure of rates and volatilities.

Further research may explore stochastic Taylor expansions with assets under stochastic volatility dynamics. The change of measure may still lead to explicit formulae for models with a known characteristic function.



\bibliographystyle{plainnat}
\bibliography{asian_expansion}

\begin{thebibliography}{15}
\providecommand{\natexlab}[1]{#1}
\providecommand{\url}[1]{\texttt{#1}}
\expandafter\ifx\csname urlstyle\endcsname\relax
  \providecommand{\doi}[1]{doi: #1}\else
  \providecommand{\doi}{doi: \begingroup \urlstyle{rm}\Url}\fi

\bibitem[Bjerksund and Stensland(2014)]{bjerksund2014closed}
Petter Bjerksund and Gunnar Stensland.
\newblock Closed form spread option valuation.
\newblock \emph{Quantitative Finance}, 14\penalty0 (10):\penalty0 1785--1794, 2014.

\bibitem[Borovkova et~al.(2007)Borovkova, Permana, and Weide]{borovkova2007closed}
Svetlana Borovkova, Ferry~J Permana, and H~vd Weide.
\newblock A closed form approach to the valuation and hedging of basket and spread options.
\newblock \emph{Journal of Derivatives}, 14\penalty0 (4):\penalty0 8, 2007.

\bibitem[Castellacci and Siclari(2003)]{castellacci2003asian}
Giuseppe Castellacci and Michael Siclari.
\newblock Asian basket spreads and other exotic averaging options.
\newblock \emph{Energy Power Risk Management}, 2003.

\bibitem[Deelstra et~al.(2010)Deelstra, Petkovic, and Vanmaele]{deelstra2010pricing}
Griselda Deelstra, Alexandre Petkovic, and Michele Vanmaele.
\newblock Pricing and hedging asian basket spread options.
\newblock \emph{Journal of computational and applied mathematics}, 233\penalty0 (11):\penalty0 2814--2830, 2010.

\bibitem[Etor{\'e} and Gobet(2012)]{etore2012stochastic}
Pierre Etor{\'e} and Emmanuel Gobet.
\newblock Stochastic expansion for the pricing of call options with discrete dividends.
\newblock \emph{Applied Mathematical Finance}, 19\penalty0 (3):\penalty0 233--264, 2012.

\bibitem[Gentle(1993)]{gentle1993basket}
David Gentle.
\newblock Basket weaving.
\newblock \emph{Risk}, 6\penalty0 (6):\penalty0 51--52, 1993.

\bibitem[Kirk and Aron(1995)]{kirk1995correlation}
Ewan Kirk and J~Aron.
\newblock Correlation in the energy markets.
\newblock \emph{Managing energy price risk}, 1:\penalty0 71--78, 1995.

\bibitem[Krekel(2003)]{krekel2003pricing}
Martin Krekel.
\newblock The pricing of asian options on average spot with average strike.
\newblock \emph{Wilmott}, \penalty0 (2004):\penalty0 80--85, 2003.

\bibitem[{Le Floc'h}(2024)]{lefloc2024stochastic}
Fabien {Le Floc'h}.
\newblock Stochastic expansion for the pricing of {Asian} and basket options.
\newblock \emph{arXiv preprint arXiv:2402.17684}, 2024.

\bibitem[Levy(1992)]{levy1992pricing}
Edmond Levy.
\newblock Pricing {European} average rate currency options.
\newblock \emph{Journal of International Money and Finance}, 11\penalty0 (5):\penalty0 474--491, 1992.

\bibitem[L’Ecuyer and Lemieux(2002)]{l2002recent}
Pierre L’Ecuyer and Christiane Lemieux.
\newblock Recent advances in randomized quasi-monte carlo methods.
\newblock \emph{Modeling uncertainty: An examination of stochastic theory, methods, and applications}, pages 419--474, 2002.

\bibitem[Margrabe(1978)]{margrabe1978value}
William Margrabe.
\newblock The value of an option to exchange one asset for another.
\newblock \emph{The journal of finance}, 33\penalty0 (1):\penalty0 177--186, 1978.

\bibitem[Pearson(1999)]{pearson1999efficient}
Neil~D Pearson.
\newblock An efficient approach for pricing spread options.
\newblock \emph{Available at SSRN 7010}, 1999.

\bibitem[Pellegrino(2016)]{pellegrino2016general}
Tommaso Pellegrino.
\newblock A general closed form approximation pricing formula for basket and multi-asset spread options.
\newblock \emph{Journal of Mathematical Finance}, 6\penalty0 (5):\penalty0 944--974, 2016.

\bibitem[Vorst(1992)]{vorst1992prices}
Ton Vorst.
\newblock Prices and hedge ratios of average exchange rate options.
\newblock \emph{International Review of Financial Analysis}, 1\penalty0 (3):\penalty0 179--193, 1992.

\end{thebibliography}

\appendix
\section{Black-Scholes identities}
Let  $\Black(F,K,v,T)$ denotes the Black-76 formula applied to the forward $F$, strike $K$ and total variance $v$ to maturity $T$:
\begin{equation}\Black(F,K,v,T)=\eta B(0,T) \left[F\Phi(\eta d_1)-K\Phi(\eta d_2)\right]\label{eqn:black}
\end{equation} with $d_1 = \frac{1}{\sqrt{v}}\left(\ln\frac{F}{K} + \frac{1}{2}v\right)$, $d_2 = d_1 -  \sqrt{v}$.

We also define \begin{equation}
	h(x) = \max(\eta x, 0)
\end{equation}
which constitutes the basis of our Taylor expansions.

\begin{identity}\label{identity:EGn}
	Let $G_n^\star = \frac{\beta}{G(T,1,m_p-1)}$ and $G_p^\star = \alpha G(T,m_p,m)$, we have
	\begin{equation}\mathbb{E}\left[ B(0,T) {G_n}^\star(T) h\left( \frac{{G_p}^\star(T)}{{G_n}^\star(T)}- \kappa^\star \right) \right] = \Black\left(1, \kappa^\star, {\nu}^2, T\right)\,,\end{equation}
	where  $\nu = \tilde{\nu}$ for the geometric approximation, $\nu$ computed from $a_i$ for the Levy approximation. 
\end{identity}
\begin{proof}
We have  $\mathbb{E}\left[ G_n^\star(T)\right] = 1$ and $G_n^\star$ may be thus expressed as \begin{equation*} G_n^\star(T)=e^{-\frac{1}{2}\int_0^T\gamma_n(s)^2\diff s + \int_0^T \gamma_n(s)  dW_{G_n}(T)}\end{equation*} in probability, with $W_{G_n}$ being a Brownian motion and $\gamma_n^2(s)=\frac{\partial \nu_n^2(s)}{\partial s}$. We interpret $\frac{G_n^\star(T)}{G_n^\star(0)}$ as a change of measure on $\mathcal{F}_T$. Under the new induced measure $\mathbb{Q}^{G_n}$, $\tilde{W}_{G_n} (T)= W_{G_n}(T) - \int_0^T \gamma_n(s) \diff s $ is a Brownian motion. Then $G_n^\star(T)$ under $\mathbb{Q}^{G_n}$ has the same law as $G_n^\star(T) e^{\nu_n^2 (T)}$ under $\mathbb{Q}$. Thus,
\begin{align}	
	\mathbb{E}\left[  {G_n}^\star(T) h\left( \frac{{G_p}^\star(T)}{{G_n}^\star(T)}- \kappa^\star \right) \right] &= \mathbb{E}^{G_n}\left[ \left( \frac{{G_p}^\star(T)}{G_n^\star(T)} e^{\nu_{np}^2(T) -\nu_n^2 (T)} - \kappa^\star\right)^+ \right]\,,\label{eqn:margrabe}
\end{align}
where 
\begin{equation}
	\nu_{np}^2(T) = -\sum_{i=1}^{m_p-1}\sum_{j=m_p}^m  a_i a_j \rho_{i,j}\int_0^T \sigma_i(s) \sigma_j(s) \diff s\,. \label{eqn:nunp}
\end{equation}
We have  \begin{equation}
	\nu^2=\nu_n^2 + \nu_p^2 - 2 \nu_{np}^2\,.\label{eqn:nu_decomp}
\end{equation}
We know that ${G_p}^\star / G_n^\star$ has variance $\nu^2$. Furthermore, $G_n^\star$ and $G_p^\star$ are martingales of unit expectation under $\mathbb{Q}$, thus \begin{equation}
\mathbb{E}\left[\frac{{G_p}^\star}{{G_n}^\star}(T)\right]=e^{-\frac{\nu_p^2(T)-\nu_n^2(T)-\nu^2(T)}{2}}\,.\label{eqn:EGpoverGn}
\end{equation}

Then using Equations \ref{eqn:nu_decomp} and \ref{eqn:margrabe}, we obtain the result.
\end{proof} 

\begin{identity}\label{identity:EGp}
	\begin{equation}
		\mathbb{E}\left[B(0,T)  G_p^\star(T) h\left( \frac{G_p^\star(T)}{G_n^\star(T)}- \kappa^\star \right)\right] = \Black\left(e^{\nu^2(T)}, \kappa^\star, \nu^2, T\right)\,.
	\end{equation}
\end{identity}
\begin{proof}
We perform a change of measure $\frac{G_p^\star(T)}{G_p^\star(0)}$. Under this new measure $\mathbb{Q}^{G_p}$, $\tilde{W}_{G_p} (T)= W_{G_p}(T) - \int_0^T \gamma_p(s) \diff s $, with $\gamma_p^2(s)=\frac{\partial \nu_p^2(s)}{\partial s}$ is a Brownian motion. Then $\frac{G_p}{G_n}^\star(T)$ under $\mathbb{Q}^{G_p}$ has the same law as $\frac{G_p}{G_n}^\star(T) e^{\nu_p^2-\nu_{pn} (T)}$ under $\mathbb{Q}$. Thus,
\begin{align}	\mathbb{E}\left[  G_p^\star(T) h\left( \frac{G_p^\star(T)}{ G_n^\star(T)}- \kappa^\star\right)\right] &= \mathbb{E}^{G_p}\left[h\left( \frac{{G_p}^\star}{G_n^\star}(T)e^{\nu_p^2(T)-\nu_{np}^2 (T)} - \kappa^\star\right) \right]\,.\label{eqn:EGpGp}
\end{align}
Using Equation \ref{eqn:nu_decomp}, we find that the right hand side corresponds to a vanilla option price under the Black model with  forward $e^{\nu^2(T)}$, strike $\kappa^\star$ and total variance $\nu^2(T)$.
\end{proof}
\begin{identity}\label{identity:ESi}
\begin{equation}
	\mathbb{E}\left[  S_{i}^{\star}(T) h\left( \frac{G_p^\star}{G_n^\star}- \kappa^\star\right)\right] = \Black\left(e^{-\frac{\nu_p^2(T)-\nu_n^2(T)-\nu^2(T)}{2}+\bar{v}_i},\kappa^\star,\nu^2,T\right)\,.
\end{equation}
with \begin{align}\bar{v}_i = \sum_{l=1}^n a_l v_{i,l}\,,&\quad v_{i,l} = \rho_{i,l}\int_{0}^T \sigma_i(s) \sigma_l(s)\diff s\,.\label{eqn:vibar_vil}\end{align}
\end{identity}
In particular, for an Asian, we have  $v_{i,l} = \int_{0}^{t_i \wedge t_j} \sigma^2(s)\diff s$. For baskets, the volatilities are taken to be constant equal to each underlying implied volatility to maturity in practice and we have  $v_{i,l} = \rho_{i,l} \sqrt{v_i v_l}$.

\begin{proof}
We use a change of measure $\frac{S_{i}^{\star}(T)}{S_i^\star(0)}$ and define \begin{equation*}\tilde{W}_i(t) = W_i(t) - \int_0^t \sigma_i(s) \diff s \end{equation*}  to obtain
\begin{align}	\mathbb{E}\left[  S_{i}^{\star}(T) h\left( \frac{G_p^\star}{G_n^\star}- \kappa^\star\right)\right] &= \mathbb{E}^i\left[h\left( \frac{{G_p}^\star}{G_n^\star}e^{ \sum_{j=1}^n a_j\int_0^{T}\rho_{i,j}\sigma_i(s)\sigma_j(s)\diff s} - \kappa^\star\right) \right]\,.\label{eqn:ESG}
\end{align}
From Equation \ref{eqn:EGpoverGn}, we find that the right hand side corresponds to a vanilla option price under the Black model with  forward $e^{-\frac{\nu_p^2(T)-\nu_n^2(T)-\nu^2(T)}{2}+\bar{v}_i}$, strike $\kappa^\star$ and total variance $\nu^2(T)$.
\end{proof} 

\begin{identity}\label{identity:EGpG}
	\begin{equation}
		\mathbb{E}\left[B(0,T)  G_p^\star(T)  \frac{G_p^\star(T)}{G_n^\star(T)}h\left( \frac{G_p^\star(T)}{G_n^\star(T)}- \kappa^\star \right)\right] = e^{\nu^2(T)}\Black\left(e^{2\nu^2(T)}, \kappa^\star, \nu^2, T\right)\,.
	\end{equation}
\end{identity}
\begin{proof}
	Let $G^\star(T) = \frac{1}{\mathbb{E}\left[\frac{G_p^\star(T)}{G_n^\star(T)}\right]} \frac{G_p^\star(T)}{G_n^\star(T)}$,
we perform a change of measure $\frac{G^\star(T)}{G^\star(0)}$. Under this new measure $\mathbb{Q}^{G}$, $\tilde{W}_{G} (T)= W_{G}(T) - \int_0^T \gamma(s) \diff s $, with $\gamma^2(s)=\frac{\partial \nu^2(s)}{\partial s}$ is a Brownian motion. Then $\frac{G_p}{G_n}^\star(T)$ under $\mathbb{Q}^{G_p}$ has the same law as $\frac{G_p}{G_n}^\star(T) e^{\nu^2(T)}$ under $\mathbb{Q}$. Thus,
\begin{multline*}	\mathbb{E}\left[  G_p^\star(T) \frac{G_p^\star(T)}{G_n^\star(T)} h\left( \frac{G_p^\star(T)}{ G_n^\star(T)}- \kappa^\star\right)\right] =\\ e^{-\frac{\nu_p^2-\nu_n^2-\nu^2(T)}{2}}\mathbb{E}^{G}\left[ G_p^\star(T) e^{\nu_p^2-\nu_{pn}^2(T)} h\left( \frac{{G_p}^\star}{G_n^\star}(T)e^{\nu^2(T)} - \kappa^\star\right) \right]\,.
\end{multline*}

Using Equation \ref{eqn:nu_decomp} and Identity \ref{identity:EGp} we find that the right hand side corresponds to a vanilla option price under the Black model with  forward $e^{2\nu^2(T)}$, strike $\kappa^\star$ and total variance $\nu^2(T)$.
\end{proof}

\begin{identity}\label{identity:ESiG}
	\begin{equation}
		\mathbb{E}\left[B(0,T)  S_i^\star(T)  \frac{G_p^\star(T)}{G_n^\star(T)}h\left( \frac{G_p^\star(T)}{G_n^\star(T)}- \kappa^\star \right)\right] = e^{\nu_n^2-\nu_{pn}^2+\bar{v}_i}\Black\left(e^{\nu^2+\nu_n^2-\nu_{pn}^2+\bar{v}_i(T)}, \kappa^\star, \nu^2, T\right)\,.
	\end{equation}
\end{identity}
\begin{proof}
	We start with a change of measure $\frac{G^\star(T)}{G^\star(0)}$: 
	\begin{multline*}
		\mathbb{E}\left[B(0,T)  S_i^\star(T)  \frac{G_p^\star(T)}{G_n^\star(T)}h\left( \frac{G_p^\star(T)}{G_n^\star(T)}- \kappa^\star \right)\right] =\\ e^{-\frac{\nu_p^2-\nu_n^2-\nu^2(T)}{2}}
		\mathbb{E}^G\left[B(0,T)  S_i^\star(T)e^{\bar{v}_i}  h\left( \frac{G_p^\star(T)}{G_n^\star(T)}e^{\nu^2 T}- \kappa^\star \right)\right]
	\end{multline*}
	We then apply the change of measure $\frac{S_{i}^{\star}(T)}{S_i^\star(0)}$:
	\begin{multline*}
		\mathbb{E}\left[B(0,T)  S_i^\star(T)  \frac{G_p^\star(T)}{G_n^\star(T)}h\left( \frac{G_p^\star(T)}{G_n^\star(T)}- \kappa^\star \right)\right] =\\ e^{-\frac{\nu_p^2-\nu_n^2-\nu^2(T)}{2}+\bar{v}_i(T)}
		\mathbb{E}^{iG}\left[B(0,T)  h\left( \frac{G_p^\star(T)}{G_n^\star(T)}e^{\nu^2 T+\bar{v}_i}- \kappa^\star \right)\right]\,.
	\end{multline*}

	Using Equations \ref{eqn:nu_decomp} and \ref{eqn:EGpoverGn}, we find that the right hand side corresponds to a vanilla option price under the Black model with  forward $e^{\frac{-\nu_p^2+\nu_n^2+3\nu^2(T)}{2}+\bar{v}_i(T)}$, strike $\kappa^\star$ and total variance $\nu^2(T)$.
\end{proof}

\begin{identity}\label{identity:ESi2overGn}
	\begin{multline}
		\mathbb{E}\left[B(0,T)  \frac{S_i^\star(T)S_j^\star(T)}{G_n^\star(T)}h\left( \frac{G_p^\star(T)}{G_n^\star(T)}- \kappa^\star \right)\right] = \\ e^{\nu_n^2+{v}_{i,j}+\bar{v}_{i}^- + \bar{v}_{j}^-(T)}\Black\left(e^{\bar{v}_{i}+\bar{v}_j(T) -2\nu_{np}^2 +2\nu_n^2(T)} , \kappa^\star, \nu^2, T\right)\,.
	\end{multline}
	with
	\begin{equation}
		\bar{v}_{i}^-(T) = \sum_{l=1}^{m_p-1} a_l\rho_{li} \int_0^T \sigma_i(s)\sigma_l(s)\diff s\,,\label{eqn:vin}
	\end{equation}	
	and $v_{i,j}$ defined in Equation \ref{eqn:vibar_vil}.
\end{identity}
\begin{proof}
We have $\mathbb{E}\left[\frac{1}{G_n^\star(T)}\right] = e^{\nu_n^2(T)}$. 
Let $G_n^{-1\star\star} =\frac{e^{-\nu_n^2(T)}}{G_n^\star(T)}$, it is a martingale of  unit expectation  under $\mathbb{Q}$ and may be thus expressed as \begin{equation*} G_n^{-1\star\star}(T)=e^{-\frac{1}{2}\int_0^T\gamma_n(s)^2\diff s + \int_0^T \gamma_n(s)  dW_{G_n^{-1}}(T)}\end{equation*} in probability, with $W_{G_n^{-1}}$ being a Brownian motion and $\gamma_n^2(s)=\frac{\partial \nu_n^2(s)}{\partial s}$. We interpret $\frac{G_n^{-1\star\star}(T)}{G_n^{-1\star\star}(0)}$ as a change of measure on $\mathcal{F}_T$. Under the new induced measure $\mathbb{Q}^{G_n^{-1}}$, $\tilde{W}_{G_n^{-1}} (T)= W_{G_n^{-1}}(T) - \int_0^T \gamma_n(s) \diff s $ is a Brownian motion. Then 
\begin{multline*}	
	\mathbb{E}\left[  \frac{S_i^\star(T)S_j^\star(T)}{G_n^\star(T)} h\left( \frac{{G_p}^\star(T)}{{G_n}^\star(T)}- \kappa^\star \right) \right] =\\ e^{\nu_n^2(T)}\mathbb{E}^{G_n^{-1}}\left[S_i^\star(T)S_j^\star(T)e^{\bar{v}_{i}^- + \bar{v}_{j}^-(T)} h\left( \frac{{G_p}^\star(T)}{G_n^\star(T)} e^{-\nu_{np}^2(T) +\nu_n^2 (T)} - \kappa^\star\right) \right]\,.
\end{multline*}
We then apply the change of measure $\frac{S_{i}^{\star}(T)}{S_i^\star(0)}$:
\begin{multline*}	
	\mathbb{E}\left[  \frac{S_i^\star(T)S_j^\star(T)}{G_n^\star(T)} h\left( \frac{{G_p}^\star(T)}{{G_n}^\star(T)}- \kappa^\star \right) \right] =\\ e^{\nu_n^2(T)}\mathbb{E}^{iG_n^{-1}}\left[S_j^\star(T)e^{{v}_{i,j}+\bar{v}_{i}^- + \bar{v}_{j}^-(T)} h\left( \frac{{G_p}^\star(T)}{G_n^\star(T)} e^{\bar{v}_{i}-\nu_{np}^2(T) +\nu_n^2 (T)} - \kappa^\star\right)\right]\,.
\end{multline*}
And finally a change of measure  $\frac{S_{j}^{\star}(T)}{S_j^\star(0)}$:
\begin{multline*}	
	\mathbb{E}\left[  \frac{S_i^\star(T)S_j^\star(T)}{G_n^\star(T)} h\left( \frac{{G_p}^\star(T)}{{G_n}^\star(T)}- \kappa^\star \right) \right] =\\ e^{\nu_n^2+{v}_{i,j}+\bar{v}_{i}^- + \bar{v}_{j}^-(T)}\mathbb{E}^{jiG_n^{-1}}\left[ h\left( \frac{{G_p}^\star(T)}{G_n^\star(T)} e^{\bar{v}_{i}+\bar{v}_j-\nu_{np}^2(T) +\nu_n^2 (T)} - \kappa^\star\right) \right]\,.
\end{multline*}
Using Equations \ref{eqn:nu_decomp} and \ref{eqn:EGpoverGn}, we find that the right hand side corresponds to a vanilla option price under the Black model with  forward $e^{\bar{v}_{i}+\bar{v}_j(T) -2\nu_{np}^2 +2\nu_n^2(T)}$. 
\end{proof}

\begin{identity}\label{identity:EGpG2}
	\begin{equation}
		\mathbb{E}\left[B(0,T)  G_p^\star(T)  \left(\frac{G_p^\star(T)}{G_n^\star(T)}\right)^2 h\left( \frac{G_p^\star(T)}{G_n^\star(T)}- \kappa^\star \right)\right] = e^{3\nu^2(T)}\Black\left(e^{3\nu^2(T)}, \kappa^\star, \nu^2, T\right)\,.
	\end{equation}
\end{identity}
\begin{proof}
	We start with the change of measure $\frac{G^\star(T)}{G^\star(0)}$
	\begin{multline*}
		\mathbb{E}\left[B(0,T)  G_p^\star(T)  \left(\frac{G_p^\star(T)}{G_n^\star(T)}\right)^2 h\left( \frac{G_p^\star(T)}{G_n^\star(T)}- \kappa^\star \right)\right] =\\ e^{-\frac{\nu_p^2-\nu_n^2-\nu^2(T)}{2}}\mathbb{E}\left[B(0,T)  G_p^\star(T)  \frac{G_p^\star(T)}{G_n^\star(T)}e^{\nu_p^2-\nu_{pn}^2+\nu^2(T)}  h\left( \frac{G_p^\star(T)}{G_n^\star(T)}e^{\nu^2(T)} - \kappa^\star \right)\right]\,.
	\end{multline*}
	We then apply Identity \ref{identity:EGpG} with the additional overall multiplicative factor \begin{equation*}e^{-\frac{\nu_p^2-\nu_n^2-\nu^2(T)}{2}}e^{\nu_p^2-\nu_{pn}^2+\nu^2(T)} =e^{2\nu^2(T)}\,,\end{equation*}
	and forward multiplier $e^{\nu^2(T)}$.
\end{proof}

\begin{identity}\label{identity:ESiG2}
	\begin{multline}
		\mathbb{E}\left[B(0,T)  S_i^\star(T)  \left(\frac{G_p^\star(T)}{G_n^\star(T)}\right)^2 h\left( \frac{G_p^\star(T)}{G_n^\star(T)}- \kappa^\star \right)\right] =\\ e^{\nu^2+2\nu_n^2-2\nu_{pn}^2+2\bar{v}_i}\Black\left(e^{2\nu^2+\nu_n^2-\nu_{pn}^2+\bar{v}_i(T)}, \kappa^\star, \nu^2, T\right)\,.
	\end{multline}
\end{identity}
\begin{proof}
	We start with a change of measure $\frac{G^\star(T)}{G^\star(0)}$: 
	\begin{multline*}
		\mathbb{E}\left[B(0,T)  S_i^\star(T)  \left(\frac{G_p^\star(T)}{G_n^\star(T)}\right)^2 h\left( \frac{G_p^\star(T)}{G_n^\star(T)}- \kappa^\star \right)\right] =\\ e^{-\frac{\nu_p^2-\nu_n^2-\nu^2(T)}{2}}
		\mathbb{E}^G\left[B(0,T)  S_i^\star(T)\frac{G_p^\star(T)}{G_n^\star(T)} e^{\bar{v}_i+\nu^2(T)}  h\left( \frac{G_p^\star(T)}{G_n^\star(T)}e^{\nu^2 T}- \kappa^\star \right)\right]
	\end{multline*}
	We then apply Identity \ref{identity:ESiG} with global multiplier 
	\begin{equation*}
		e^{-\frac{\nu_p^2-\nu_n^2-\nu^2(T)}{2}} e^{\bar{v}_i+\nu^2(T)}
	\end{equation*}
	and forward multiplier $e^{\nu^2 T}$.
\end{proof}

\begin{identity}\label{identity:ESi2GoverGn}
	\begin{multline}
		\mathbb{E}\left[B(0,T) \frac{S_{i}^{\star}S_{j}^{\star}G_p^\star(T)}{{G_n^\star}^2}		h\left( \frac{G_p^\star(T)}{G_n^\star(T)}-\kappa^\star \right)\right] = \\ e^{3\nu_n^2-2\nu_{pn}^2+\bar{v}_i+\bar{v}_j+{v}_{i,j}+\bar{v}_{i}^- + \bar{v}_{j}^-(T)}\Black\left(e^{\bar{v}_{i}+\bar{v}_j(T) +\nu^2 -2\nu_{pn}^2 +2\nu_n^2(T)} , \kappa^\star, \nu^2, T\right)\,.
	\end{multline}
\end{identity}
\begin{proof}
	We start with a change of measure $\frac{G^\star(T)}{G^\star(0)}$: 
	\begin{multline*}
		\mathbb{E}\left[B(0,T)  \frac{S_{i}^{\star}S_{j}^{\star}G_p^\star(T)}{{G_n^\star}^2}	h\left( \frac{G_p^\star(T)}{G_n^\star(T)}- \kappa^\star \right)\right] =\\ e^{-\frac{\nu_p^2-\nu_n^2-\nu^2(T)}{2}}
		\mathbb{E}^G\left[B(0,T) \frac{S_i^\star(T)S_{j}^{\star}}{G_n^\star(T)} e^{\bar{v}_i+\bar{v}_j+\nu_n^2(T)-\nu_{pn}^2}  h\left( \frac{G_p^\star(T)}{G_n^\star(T)}e^{\nu^2 T}- \kappa^\star \right)\right]
	\end{multline*}
	We then apply Identity \ref{identity:ESi2overGn} with global multiplier 
	\begin{equation*}
		e^{-\frac{\nu_p^2-\nu_n^2-\nu^2(T)}{2}} e^{\bar{v}_i+\bar{v}_j+\nu_n^2(T)-\nu_{pn}^2}
	\end{equation*}
	and forward multiplier $e^{\nu^2 T}$.
\end{proof}

\begin{identity}\label{identiy:SiSjSkoverGn2}
	\begin{multline}
		\mathbb{E}\left[B(0,T)  	\frac{S_{i}^{\star}S_{j}^{\star}S_{k}^{\star}(T)}{{G_n^\star}^2}h\left( \frac{G_p^\star(T)}{G_n^\star(T)}- \kappa^\star \right)\right] = \\ e^{{v}_{i,j}+{v}_{k,i}+{v}_{k,j}+2\bar{v}_{i}^- + 2\bar{v}_{j}^- +2\bar{v}_{k}^- + 3\nu_n^2(T)}\Black\left(e^{\bar{v}_{i}+\bar{v}_j+\bar{v}_{k} -3\nu_{np}^2 +3\nu_n^2(T)} , \kappa^\star, \nu^2, T\right)\,.
	\end{multline}
\end{identity}
\begin{proof}
We start with a change of measure $\frac{G_n^{-1\star\star}(T)}{G_n^{-1\star\star}(0)}$:
\begin{multline*}	
	\mathbb{E}\left[  \frac{S_i^\star(T)S_j^\star(T)S_{k}^{\star}(T)}{{G_n^\star(T)}^2} h\left( \frac{{G_p}^\star(T)}{{G_n}^\star(T)}- \kappa^\star \right) \right] =\\ e^{\nu_n^2(T)}\mathbb{E}^{G_n^{-1}}\left[ \frac{S_i^\star(T)S_j^\star(T)S_{k}^{\star}(T)}{{G_n^\star(T)}}e^{\bar{v}_{i}^- + \bar{v}_{j}^- + \bar{v}_{k}^-(T)+\nu_n^2 (T)} h\left( \frac{{G_p}^\star(T)}{G_n^\star(T)} e^{-\nu_{np}^2(T) +\nu_n^2 (T)} - \kappa^\star\right) \right]\,.
\end{multline*}
We then perform a change of measure $\frac{S_k^\star(T)}{S_k^\star(0)}$:
\begin{multline*}	
	\mathbb{E}\left[   \frac{S_i^\star(T)S_j^\star(T)S_{k}^{\star}(T)}{{G_n^\star(T)}^2} h\left( \frac{{G_p}^\star(T)}{{G_n}^\star(T)}- \kappa^\star \right) \right] =\\ e^{\nu_n^2(T)}\mathbb{E}^{kG_n^{-1}}\left[ \frac{S_i^\star S_j^\star(T)}{{G_n^\star(T)}}e^{{v}_{k,i}+{v}_{k,j}+\bar{v}_{i}^- + \bar{v}_{j}^- +2\bar{v}_{k}^- +\nu_n^2 (T)} h\left( \frac{{G_p}^\star(T)}{G_n^\star(T)} e^{\bar{v}_{k}-\nu_{np}^2(T) +\nu_n^2 (T)} - \kappa^\star\right)\right]\,.
\end{multline*}
And we apply Identity \ref{identity:ESi2overGn}  with global multiplier 
\begin{equation*}
	e^{{v}_{k,i}+{v}_{k,j}+\bar{v}_{i}^- + \bar{v}_{j}^- +2\bar{v}_{k}^- + 2\nu_n^2(T)}
\end{equation*}
and forward multiplier $e^{\bar{v}_{k}-\nu_{np}^2(T) +\nu_n^2 (T)}$.

\end{proof}

\section{Derivatives of the Black-76 formula towards the strike}\label{sec:appendix_black_der}
The following formulae are used to compute the expansions in practice:
\begin{align}
	\frac{\partial}{\partial K} \Black(F, K, v, T) &= -\eta B(0,T)  \Phi(\eta d_2)\,,\\
	\frac{\partial^2}{\partial K^2} \Black(F, K, v, T) &= \frac{B(0,T)}{K\sqrt{v}}  \phi(d_2)\,,\\
	\frac{\partial^3}{\partial K^3} \Black(F, K, v, T) &= \frac{ B(0,T)}{K^2\sqrt{v}}  \left( \frac{\phi( d_2)}{\sqrt{v}}-1\right)\,,
\end{align}
where $\Phi$, $\phi$ are the cumulative normal distribution function and the normal probability density function. 
 
\end{document}